
\magnification=\magstep1
\font\B=cmbx10 scaled\magstep1
\font\cs=cmcsc10
\font\eightrm=cmr8
\baselineskip=21pt
\def\frac#1#2{#1\over #2}

\def\Ntr{{\cal N}_{\hbox{\eightrm tr}}}
\def\Nrot{{\cal N}_{\hbox{\eightrm rot}}}

{}~~~~
\vskip.5truein

\centerline{\B Magnetic Screening in Thermal Yang-Mills Theories}
\vskip.5truein

\centerline{\cs T. S. Bir\'o}

\centerline{Institut f\"ur Theoretische Physik,
Justus-Liebig Universit\"at,}

\centerline{Heinrich-Buff-Ring 16, D-6300 Gie{\ss}en, Germany}

\medskip
\centerline{and}
\medskip

\centerline{\cs B. M\"uller}

\centerline{Department of Physics, Duke University, Durham, NC 27708-0305}
\vskip.5truein

\centerline{\bf Abstract}
\medskip

{\narrower
We develop a semiclassical method to calculate the density of magnetic
monopoles in non-abelian gauge theories at finite temperature in the
dilute gas approximation.  This quantity is related
to the inverse magnetic screening length for which we
obtain  $\mu_M = 0.255 g^2T$ in SU(2).
\vskip.5truein}

\noindent {\bf 1. Introduction}

The screening of static magnetic gauge fields in Yang-Mills theories
at high temperature has long proved to be intractable by analytic
techniques.  There exist well known arguments$^1$ that the inverse
screening length $\mu_M$ must be of order $g^2T$, and
this has been partially supported by numerical calculations in the
framework of lattice gauge theory$^2$ which have yielded the result
$\mu_M = (0.27 \pm 0.03) g^2T\quad \hbox{for SU(2)}$.
The recently developed resummation techniques$^3$ for the
finite-temperature gauge theories cure many infrared divergences
arising in the perturbative expansion, but not those associated with
long-range static magnetic gauge fields.  Although the Schwinger-Dyson
equation that determines $\mu_M$ has been identified some time ago$^4$,
its solution remains unknown.  A recent attempt$^5$ to calculate
magnetic screening to leading order in dual QCD has led to the result
$2\pi T / g$ for the dual gluon mass in the high-temperature limit,
which is not directly related to the static magnetic screening mass.

Here we propose a new approach to the calculation of $\mu_M$, which is
not based on a perturbative expansion in the gauge coupling constant
$g$.  Our starting point is the observation\footnote{$^*$}{Note that
in our convention $g$ is defined by the {\it classical} Yang-Mills
equations, and is related to the dimensionless coupling constant by
$\alpha_s = g^2\hbar/4\pi$.}  that the
combination $g^2T$ defines an inverse length without any factor involving
powers of Planck's constant $\hbar$.  One may therefore speculate that
$\mu_M$ can be calculated from {\it classical} Yang-Mills statistical
mechanics, with quantum effects providing corrections of higher order
in $g$.  This conjecture was recently found to be true for the thermal
gauge boson (plasmon) damping rate, which also is of order $g^2T$
and can be calculated either by resummation techniques$^6$ or by a
stability analysis of time-dependent classical gauge field
configurations$^7$. This coincidence is nontrivial, because in
the perturbative calculation the gluon damping rate is seen to
superficially depend on the electric screening mass, which is of
order $gT/\sqrt{\hbar}$. However, this dependence exactly cancels
in the leading order result.

Similarly, it has been emphasized by Landsman$^8$ that the effective
high-temperature effective action for the static sector
of gauge theories contains quantum corrections describing the
Debye screening of static electric fields. These effectively convert
the dimensionally reduced Yang-Mills action into a Yang-Mills--Higgs
action that remains unbroken at tree level. Magnetic monopoles can
induce a symmetry breaking term into the effective Higgs action, as
pointed out by Oleszczuk and Polonyi$^9$. Our treatment assumes
that these effects of the full quantum theory do not contribute to the
magnetic screening mass at leading order in $g$. Whether that is
indeed so will require futher investigation.

This paper is organized as follows:  we connect the magnetic screening
mass in high-temperature Yang-Mills theories with the magnetic
monopole charge density in the framework of the linear response
theory.  We then calculate the density of magnetic monopole charges
from the canonical partition function in Chapter 3.  A saddle point
evaluation in the space of general radial symmetric monopole field
configurations becomes possible by inserting an appropriate unit
factor, which sets the scale and stabilizes classical soliton
solutions of a given size.  The integration over scale sizes is done
at the end.  Besides these stabilized classical field configurations
the contributions of fluctuation modes with zero frequency, i.e. the
collective modes of translation and rotation, must be taken into
account because they determine the entropy of a given field
configuration.  Finally, in Chapter 4, we conclude by exploring the
characteristic size and energy of dominant monopole configurations
under conditions that may be reached in heavy-ion collisions at RHIC
energies.
\bigskip

\noindent {\bf 2. Magnetic Screening}
\medskip

In order to obtain the magnetic screening length one usually
investigates the linear response of the medium to an infinitesimal
external magnetic field.  The magnetic field, which couples to the
monopole charge, is curl-free and can be expressed by a scalar
magnetic potential $\phi^a$,
$$B_i^a = -\partial_i \phi^a. \eqno(1)$$
The application of such an external magnetic field modifies the
monopole charge density in the plasma, so the partition function becomes
$$Z(\phi)= \hbox{Tr}\; e^{-\beta(\hat H-\phi^a\hat Q^a)}, \eqno(2)$$
where $\hat Q^a$ is the magnetic charge operator and the trace ``Tr''
runs over all possible states of the Yang-Mills field.
Here we discuss how to evaluate this partition function in the case
of SU(2) only, but the result can easily be generalized to SU(N).
Let us denote a given color charge state by $\vert q, q_3\rangle$,
where  $q$, the eigenvalue of the Casimir operator, is a multiplet
index. A magnetic monopole state corresponds to the eigenvalue $q=1$.

Since the Yang-Mills action is color symmetric the eigenstates of the
Hamiltonian are $q_3$ independent, so one can always ``rotate'' an
energy and color eigenstate $\vert \omega, q_3\rangle$ so that the action
of $\beta \phi^a \hat Q^a$ can be represented by that of the generator
of the abelian subgroup $\Theta \hat I_3 \equiv g^{-1}\beta\vert\phi\vert
\hat I_3$, where $g^{-1}$ is the elementary magnetic charge,
$\hat I_3$ has the eigenvalues $q_3 = -1$, $0$ and $+1$ and finally
$$ \vert \phi \vert^2 = \phi^a \phi^a. \eqno(3)$$
Using this particular representation of gluon states $\hat I_3$ and
$\hat H$ commute and in the dilute gas approximation the quantum
numbers $\omega$ and $q_3$ are additive, so the partition function (2)
factorizes for different $\omega$ and $q_3$ values.$^{10}$
Taking into account the bosonic nature of excitations of the
Yang-Mills field, the magnetic monopole partition function in the
dilute gas limit becomes:
$$Z = \prod_{\omega} \prod_{q_3} \sum_{n=0}^{\infty}
e^{-n(\beta \omega - q_3 \Theta)}, \eqno(4)$$
where $q_3$ runs over the possibilities $-1$, $0$ and $+1$.

The summation over the indefinite occupation number $n$ yields a
Bose-Einstein factor for each $\omega$ and $q_3$.
{}From this partition function we obtain the magnetic color charge
density using the standard formula
$$ \rho^a = {1 \over \beta V} {\partial \over \partial \phi^a} \ln Z.
\eqno(5)$$
In the weak external field limit ($\Theta \ll 1$) this leads to
(see Appendix A for details):
$$ \rho^a = - {N \over Vg^2T} Z_M \phi^a \eqno(6)$$
for the group SU(N) with the effective one-monopole partition function
$$Z_M = \sum_{\omega} {1 \over e^{\beta \omega}-1}. \eqno(7) $$
Because of the large average mass of a monopole
we can make use of the Boltzmann approximation, obtaining
$$ Z_M \approx \sum_{\omega} e^{-\beta \omega} = Tr \left( e^{- \beta
{\hat H}} \right). \eqno(8) $$

Since the magnetic monopole charge density is the source of the
divergence of the magnetic field
$$\partial_i B_i^a = \rho^a, \eqno(9)$$
we arrive at the following equation describing the linear
polarizability of a magnetic monopole gas
$$\partial_i\partial_i \phi^a - {N Z_M\over g^2 T V}
\phi^a = 0. \eqno(10)$$
Inspecting this equation one easily realizes that an effective
magnetic screening mass is obtained,
$$\mu_M = \left( {N Z_M\over (g^2T)^3V}\right)^{{1\over 2}} g^2T,
\eqno(11)$$
causing an exponential damping of the external magnetic potential
$$\phi^a(r) \sim e^{-\mu_M r}. \eqno(12)$$

It is reasonable to expect that the ratio $Z_M/V$ scales with $T^3$
at high temperatures because $T$ is the only available scale.  Recall
furthermore that the four-dimensional gauge theory at high temperatures
undergoes dimensional reduction and becomes equivalent to a
three-dimensional gauge theory with the effective temperature $g^2T$.
We therefore expect that the partition
function scales as $Z_M \sim V (g^2 T)^3$, so the magnetic
screening mass is proportional to $g^2T$.

It is illuminating to repeat the same considerations for the electric
sector of the gauge theory. In this case the external potential must
be taken as that of a static color-electric field, and one considers
the dilute gas of gauge field excitations with electric charge in the
adjoint representation. In lowest perturbative order these are just
the free perturbative gauge bosons, but it is now inappropriate to
use the Boltzmann approximation because there is no mass gap.
As shown in Appendix A, this consideration yields precisely the
electric screening mass $\sqrt{N/3}gT$ obtained in diagrammatic
perturbation theory.
\bigskip

\noindent {\bf 3. Partition Function of Magnetic Monopoles}
\bigskip

\noindent {\cs 3.1. Spherically Symmetric Gauge Fields}
\medskip

In order to calculate the nontrivial factor we need to obtain the
partition function $Z_M$.  Although
simple scaling arguments$^{1,11}$ show that there may be a
characteristic monopole size $R_0$ contributing dominantly to the
partition sum
$$Z_M \sim \int {dR\over R^5} \exp(-R_0/R), \eqno(13)$$
no classical monopole solution exists, which would offer a stable
stationary point of the path integral defining the partition function.

While the t'Hooft-Polyakov solution$^{12}$ for spontaneously broken SU(2)
gauge theories is stable and has finite energy, its counterpart in
pure SU(2) gauge theory, the Wu-Yang monopole$^{13}$, has infinite energy
and is unstable against small perturbations$^{14,15}$.  This lack of a basis
for a semiclassical analysis of the statistical mechanics of monopole
solutions in non-abelian gauge theories apparently vitiates an analytical
approach.  Here we want to sketch how the stability problem may be
circumvented, opening the way to a semiclassical calculation of the magnetic
monopole density at high temperature, and therefore of the magnetic
screening length.

Let us begin by considering SU(2) gauge field configurations which carry
one unit of magnetic charge.  They must asymptotically look like a
monopole $(j=0)$ mode of the operator
$$\hat J = \hat S + \hat I + \hat L,\eqno(14)$$
where $\hat S$ and $\hat I$ denote the generators of spin and color spin
in the adjoint representation and $\hat L$ is the generator of orbital
angular momentum for the gauge field.  Since the gauge field belongs
to the representation $S=I=1$, there are three different possibilities to
construct a $j=0$ mode; namely $\vert L,T\rangle = \vert 1,1\rangle,\;
\vert 0,0\rangle$, and $\vert 2,2\rangle$ combinations, where the
grand-spin quantum number $T$ is obtained from the eigenvalue of the
Casimir operator ${\hat T}^2=T(T+1)$ with
$$\hat T = \hat S + \hat I.\eqno(15)$$
The Wu-Yang monopole, known to be unstable, belongs to the $j=0$ mode
of type $\vert L,T\rangle = \vert 1,1\rangle$. A further unstable mode
has been found to involve a combination of the two other states by Akiba,
Kikuchi and Yamagida$^{16}$.

Having this in mind we start our investigation with the most general
ansatz$^{17}$ for the monopole vector potential
$$A_{ia} \equiv \sum_{\alpha=\pm,0} P_{ia}^{(\alpha)}A^{(\alpha)}
= {1\over r} \left( P_{ia}^+ (ue^{i\phi}-i)+
P_{ia}^-(ue^{-i\phi}+i) + P_{ia}^0 w\right), \eqno(16)$$
where $u(r),\;\phi(r)$ and $w(r)$ are real functions of the radial
variable only and we use the projectors
$$\eqalignno{P_{ia}^{\pm} &= \textstyle{{1\over 2}} (\delta_{ia} -
n_in_a \pm i\epsilon_{iaj}n_j), \cr
P_{ia}^0 &= n_in_a &(17) \cr} $$
with the unit radial vector $n_i=x_i/r$.  Configurations of the
Wu-Yang monopole-type are proportional to ${1\over 2i}(P_{ia}^+ -
P_{ia}^-) = \epsilon_{iaj}n_j$, i.e. they belong to the choice $w=0$ and
$u=0$.  The unstable mode found by Akiba et al. is proportional to a
linear combination of ${1\over 2}(P_{ia}^+ + P_{ia}^-)$ and $P_{ia}^0$,
hence corresponds to $w\not= 0$ and $\sin\phi=0$.

The different components of this ansatz, $A^+, A^-$ and $A^0$, the
respective coefficients of the projectors, are related to the $\vert
L, T\rangle$ specification of the monopole mode $(j=0)$ by simple
linear combinations
$$\vert 0,0\rangle = {2\over 3} (A^++A^-) + {1\over 3}
A^0, \eqno(18)$$
\medskip
$$\vert 1,1\rangle = {i\over 2} (A^+-A^-) \eqno(19)$$
and
$$\vert 2,2\rangle = {1\over 3} (A^++A^-) - {1\over 3}A^0. \eqno(20)$$
The magnetic field described by our ansatz is
$$B_{ia} = {1\over r^2} \left[ P_{ia}^+ \left(iru' +u(r\phi'+w)
\right) e^{-i\phi} + c.c. + P_{ia}^0(1-u^2)\right] \eqno(21)$$
where the prime denotes radial differentiation and $c.c.$ stands for
complex conjugate.

The magnetic monopole charge seen from outside a sphere of radius $r$
can be obtained from the magnetic analogue of Gauss' law
$$Q_a(r) = {1\over 4\pi}\int d^3r \partial_iB_{ia} =
{1\over 4\pi}\oint_r n_iB_{ia} = n_a(1-u^2(r)). \eqno(22)$$
This result shows that a monopole field configuration requires
asymptotically $u\to 0$ as $r\to \infty$, irrespective to the fields
$w(r)$ and $\phi(r)$.

The energy of a static configuration in the high-$T$ limit defines the
effective action of the dimensionally reduced euclidean field theory
$$S_3[A] = \beta E = {1\over g^2T}\int d^3r {1\over 2}
(E_{ia}E_{ia}+B_{ia}B_{ia}). \eqno(23)$$
Introducing the scaled parameter $\tilde{\beta} = {4\pi\over g^2T}$, we
find for the ansatz (16):
$$\eqalignno{\beta E &= \tilde{\beta} \int_0^{\infty} dr \left[ \left(
{du\over dr}\right)^2 + u^2 \left( {d\phi\over dr} + {1\over r}w \right)^2
+ {(1-u^2)^2\over 2r^2}\right]\cr \cr
&= \tilde{\beta} E[u] + \tilde{\beta} \int_0^{\infty} dr\; u^2
\left({d\phi\over dr} + {w\over r}\right)^2. &(24) \cr}$$

Here one observes that only the field $u(r)$ is really a dynamical
degree of freedom while $w(r)$ is non-dynamical and $\phi(r)$ is a
cyclic variable.  Their physical interpretation becomes clear inspecting
infinitesimal gauge transformations of the vector potential
$$\delta A_{ia} = D_{iab}\delta\Lambda_b, \eqno(25)$$
where
$$D_{iab} = \delta_{ab}\partial_i - \epsilon_{acb}A_{ic} \eqno(26)$$
is the gauge-covariant derivative.  Any gauge transformation which
conserves the monopole form  of our ansatz (16) must have the form
$$\delta\Lambda_b = n_b\cdot \delta\Lambda(r). \eqno(27)$$
This implies that restricting ourselves to static magnetic monopole
gauge field configurations there still remains a residual gauge degree of
freedom.  The variation of the ansatz fields under such an infinitesimal
gauge transformation is given by
$$\eqalignno{\delta u &= 0,\cr \delta\phi &= -\delta\Lambda,\cr
\delta w &= r {d\over dr} \delta\Lambda. &(28) \cr}$$
The field $u(r)$ and hence the monopole charge is unchanged, but the
field $\phi$ is rotated by a gauge transformation mixing fluctuations of
the pure Wu-Yang ansatz with the other unstable mode.  The meaning of the
field $w(r)$ is less obvious, we note only that it vanishes in the
Schwinger gauge $x_iA_{ia} = 0$.
\bigskip

\noindent {\cs 3.2. The Functional Integral}
\medskip

Inspecting the form of the effective action $S_3$, one realizes that
the path integral in the canonical partition sum can be easily done in two
of the fields, $w(r)$ and $\phi(r)$.  Turning from the `Cartesian' field
variables $A_{ia}$ (in mode $j=0$) to the `cylindrical' variables $u,
\phi$ and $w$ introduced before we use the integration measure
$${\cal D}A_{ia}^{j=0} = {u\over r^3}  {\cal D}u\;{\cal D}\phi\;
{\cal D}w. \eqno(29)$$
Because the ansatz (16) is invariant under transformation by $\hat
J$, the single monopole partition sum factorizes for small
fluctuations around the monopole form:
$$Z_M = \int {\cal D}Ae^{-S_3[A]} = \int {\cal D}A^{j=0}\;
e^{-\beta E[u,\phi,w]} \prod_{j>0} Z_j[u] \eqno(30)$$
where
$$Z_j[u] = \int {\cal D}(A^j) \exp\left( -S_3\left[ A^{j=0}
+\delta A^j\right] + S_3\left[ A^{j=0}\right] \right) \eqno(31)$$
depends only on $u(r)$ due to the gauge freedom (28).  Here we have
anticipated that we will find a nontrivial stationary point in the
mode $u(r)$, around which we can expand the functional integration.
All integrations in (31) will be Gaussian except those corresponding
to the zero-modes associated with translations and rotations:
$$Z_{j>0} \equiv \prod_{j>0} Z_j[u] = Z_{\hbox{\eightrm tr}} [u]
Z_{\hbox{\eightrm rot}}[u] \prod_{j>0}{'}  Z_j[u], \eqno(32)$$
where the prime indicates that the zero-modes have been separated.
{}From studies of small fluctuations around the sphaleron$^{18,19}$ in
spontaneously broken SU(2) gauge theory it is known that all other
modes are stable, and lead to well-behaved Gaussian integrals in
the semiclassical approximation.

The Gaussian integral in the variable $w$ can be easily done.
After doing that we get
$$\int {u\over r} {\cal D} w\; \exp\left( -\tilde{\beta}
\int_0^{\infty} dr \bigg( {w\over r} + {d\phi\over dr}\bigg)^2
u^2\right) = \hbox{const.} . \eqno(33)$$
compensating for the factor $u$ in (29). We normalize the constant
to be unity.
Now the compact integral over the field $\phi$ is trivial and can
be normalized such that
$$\int{\cal D}\phi = 1, \eqno(34)$$
resulting in a functional integral over $u(r)$
$$Z_M = \int {\cal D} u\; e^{-\tilde{\beta}E[u]} Z_{\hbox{\eightrm
tr}}[u] Z_{\hbox{\eightrm rot}} [u], \eqno(35)$$
if we neglect the influence of non-collective multipole $(j>0)$ fluctuations
on the partition function. This issue will be briefly discussed at the
end of this Chapter again.

One may wonder at this point why we have chosen these particular
normalizations.  The physical idea behind it is that for free fields
one has to arrive at the canonical partition sum at high temperature.
Since the only dynamical degree of freedom is represented by $u$, the
other two fields must not contribute to a correctly normalized path
integral.
\bigskip

\noindent {\cs 3.3. Setting the Scale}
\medskip

Unfortunately, $E[u]$ does not have a stable saddle point $u_0(r)$.
This lack of a basis for the semiclassical expansion, however, does
not preclude the calculation of the partition function for gauge field
configurations $u$ with monopole symmetry.  It only implies that the
evaluation of the functional integral (35) cannot be restricted to
integration over Gaussian fluctuations around a classical solution
$u_0(r)$.  We now propose a method how this functional integral (35),
where $u(r)$ satisfies the boundary conditions $u(\infty)=0,\; u(0)=1$,
may be calculated.  It may be practically carried out by adding a
stabilizing term to the expression for $E[u]$,
$$E[u] = \int_0^{\infty} dr \left[ \left( {du\over dr}\right)^2 +
{(1-u^2)^2\over 2r^2}\right], \eqno(36)$$
introduced in eq. (24).  The idea is to introduce a length scale that
favors monopole configurations of a particular core size, and then
integrate over the dummy scale in such a way that the partition function
remains unchanged.  Inspired by the analogous expression for the
t'Hooft-Polyakov monopole$^{12}$, where a stabilizing mass term
$(g\Phi)^2u^2$ involving the Higgs field $\Phi$ appears, we introduce
the term $\Delta E[u] = \lambda^2 D[u]$,
$$D[u] = {1\over 2} \int_0^{\infty} dr \left( 1-(1-u^2)^2\right) ,
\eqno(37)$$
by inserting a unit factor
$$1= \int_0^{\infty} d(\lambda^2) \; \tilde{\beta}
D[u]\; e^{-\tilde{\beta}\lambda^2 D[u]} \eqno(38)$$
into the integral (35).  Here $\lambda$ has the dimension of an
inverse length.  At large $r$, in view of the boundary condition $u(r)
\buildrel {r\to\infty}\over\longrightarrow 0$, it acts like a mass
term leading to the asymptotic solution
$u(r) \to A\exp (-\lambda r)$,
while it does not interfere with the limit $u(0) = \pm 1$.
Our choice of $D[u]$ is unique\footnote{$^*$}{We have explored
the consequences of the modified scale breaking factors
$D[u] = {1\over N} \int dr(1-(1-u^2)^N)$, which are of higher order
in $u(r)$.  For $N=3,4$ these yield somewhat smaller values
(by 7 and 14 percent, respectively) for the monopole density within
our approximation. This indicates that our result is not very
sensitive to the precise form of the scale breaking term. We note
that the functional integral over $u(r)$ in (35) could, of course,
be performed numerically by Monte-Carlo integration, avoiding the
errors introduced in the saddle-point approximation.}
if one requests that the integrand be an even function of not higher
than fourth order in $u$ which leads to a vanishing energy density at
$r\to \infty$.

After inserting the unit factor (38) into the functional integral (35)
we interchange the order of integration over the dummy scale
parameter, $\lambda$, and with that over the fields.  A consistent use
of the dimensionless variable $x=\lambda r$ leads now to the following
expression:
$$Z_M = \int_0^{\infty} d\lambda  \; 2\tilde{\beta}
\int {\cal D} u\; D[u] e^{-\tilde{\beta}(E[u]+\Delta E[u])}
Z_{\hbox{\eightrm tr}} Z_{\hbox{\eightrm rot}}, \eqno(39)$$
where
$$E[u] = \lambda \int_0^{\infty} dx \left[ \left( {du\over
dx}\right)^2 + {(1-u^2)\over x^2}\right], \eqno(40)$$
$$D[u] = {1\over{2\lambda}} \int_0^{\infty} dx (2u^2-u^4), \eqno(41)$$
$$\Delta E[u] = \lambda^2 D[u]. \eqno(42)$$
The functional integral over $u$ in (39)  can now be approximated by a
Gaussian integration around the lowest energy stationary solution
$u(r)=u_0(x)$ of the exponent, which satisfies the equation
$${d^2u_0\over d x^2} + u_0 (1-u_0^2) \left({1\over x^2}
-1\right) = 0. \eqno(43)$$
Since there is no other scale involved besides the dummy parameter
$\lambda$, the solution of eq. (43) is solely a function of the
dimensionless variable $x$, and the ground state energy scales
as
$$E[u_0] + \Delta E[u_0] = \lambda a. \eqno(44)$$
By numerical integration of (43) we have found the value $a= 1.469$.
The function $u_0(r)$ is displayed in Figure 1.  We also obtain
$\Delta E[u_0] = b\lambda$ with $b= 0.695$.  The monopole charge
contained inside the radius $r$ (eq. 22) is also shown in Fig. 1.  The
radius $R_{1/2}$ inside which half of the asymptotic charge is contained
can be defined as the size of the monopole-soliton.  This happens to be
the inflection point of the solution $u_0(\lambda r)$
$$R_{1/2} = \lambda^{-1}.\eqno(45)$$
Postponing the Gaussian integration over small fluctuations around
$u_0$ we are left with a single-parameter integral:
$$Z_M \approx 2 \tilde{\beta} \int_0^{\infty} d\lambda \;
b \;e^{-\tilde{\beta} a\lambda}\; Z_{\hbox{\eightrm tr}}
Z_{\hbox{\eightrm rot}}. \eqno(46)$$
\bigskip

\noindent {\cs 3.4. Zero-Mode Contributions}
\medskip

Now we turn to the determination of the zero-mode contributions to
$Z_M$.  The method we use is that of collective coordinates,$^{20}$ which
leads to an integral over the respective group volumes of translations
and rotations, $V$ and $8\pi^2$.  What remains is to take into account
the normalization of the wave functions describing these modes.
Physically they are related to the total momentum (energy) of a
monopole soliton and to its rotational inertia, respectively, as
$Z_{\hbox{\eightrm tr}} = \lambda^3 V \Ntr^3$,
since the classical solution is a function of $x = \lambda r$, and
$Z_{\hbox{\eightrm rot}}=8\pi^2 \Nrot^3$, with
$$\eqalignno{\Ntr^2 &= {1\over 6\pi} \int d^3r\; \epsilon(r), &(47)\cr
\Nrot^2 &= {1\over 6\pi} \int d^3r\left( r^2\epsilon (r) -
\hbox{correction}\right), &(48)\cr}$$
where $\epsilon(r)$ is the energy density of the classical soliton,
and the correction---due to an extra gauge rotation---ensures the
correct boundary condition for the rotational energy density at radial
infinity (see ref. 19). The collective wave function would
have the form
$$\varphi = {1\over\sqrt{2E}}\; e^{iP\cdot x}\; f, \eqno(44)$$
with a form factor $f$ normalized so that
including a spin factor $3$ for massive solitons and a radial normalization
factor $4\pi$ we have
$$\vert\varphi\vert^2 = \Ntr^{-2} \cdot {1\over 2E} \cdot 3\cdot 4\pi = 1,
\eqno(50)$$
whence we obtain eq.(47).  While the total energy of the soliton scales
like $E=4\pi\lambda a$, the rotational inertia of a homogeneous sphere
with radius $R$ like ${3\over 5} E R^2 = {3\over 5}4\pi a\big/\lambda$,
because the characteristic radius of the soliton is
$\lambda^{-1}$. This yields the approximate expressions
$$\eqalignno{\Ntr^2 &\approx {2\over 3}\lambda a, &(51)\cr
\noalign{\hbox{and}}
\Nrot^2 &\approx {2\over 5} {a\over\lambda} , &(52)\cr}$$
which lead to the zero-mode factors
$$\eqalignno{Z_{\hbox{\eightrm tr}} &= \lambda^3
\overline{Z}_{\hbox{\eightrm tr}}  = \left( {2\over 3}a\right)^{{3\over 2}}
\;\lambda^{{3\over 2}} \; \lambda^3 V, &(53) \cr
Z_{\hbox{\eightrm rot}} &= \left( {2\over 5} a\right)^{{3\over 2}}\;
\lambda^{-{3\over 2}}\; 8\pi^2. &(54) \cr}$$
We note that their product scales like
$\lambda^3$.  The integral over $\lambda$ in (46) can now be carried
out, yielding
$$Z_M = 12 \overline{Z}_{\hbox{\eightrm tr}} Z_{\hbox{\eightrm rot}}
{b\over \tilde{\beta}^3 a^4}. \eqno(55)$$
Inserting (53,54) we finally obtain the monopole density in the classical
limit as
$$\rho_M = {1\over V}Z_M = {12b\over 15^{3/2}\pi a} (g^2T)^3
= 0.0657\; {b\over a}\; (g^2T)^3. \eqno(56)$$
\bigskip

\noindent {\cs 3.5. Radial Monopole Fluctuations}
\medskip

We finally evaluate the scaled determinantal contribution,
$\Delta^{-1/2}$, due to the fluctuations around the soliton-monopole
solution $u_0(r)$:
$$\Delta = \prod_n \omega_n^2. \eqno(57)$$
Here $\omega_n^2$ are the eigenvalues of the operator
$$\Omega^2 = -{d^2\over dx^2} + (1-3u_0^2(x)) \left( 1-{1\over x^2}\right)
\eqno(58)$$
obtained expanding the effective action $E[u]+\Delta E[u]$ up to second
order in $(u-u_0)$.  The effective potential involved in this eigenvalue
problem,
$$V_{\hbox{\sevenrm eff}}(x) = \left(1-3u_0^2(x)\right)\cdot
\left( 1-{1\over x^2}\right) \eqno(59)$$
is plotted in Figure 2.  Since we are calculating the partition sum
$Z_M$, which is restricted to single monopole configurations, we only take
into account fluctuations around $u_0$ that vanish at infinity.  Hence
only those eigenvalues corresponding to a bound state in this effective
potential contribute, i.e.
$$\Delta = \prod_n \omega_n^2\; \theta(1-\omega_n^2) \eqno(60)$$
where $\theta$ is the step function.  Numerically we have found only
one such bound state with the eigenvalue $\omega_0^2 = 0.950$.

However, the asymptotic form of $V_{\hbox{\eightrm eff}}$, $-{1\over
r^2}$, is known to have infinitely many bound states close to threshold.
Their eigenvalues, estimated from the Bohr-Sommerfeld formula, are
$$\omega_n^2 = \left( 1-e^{-(2n+1)\pi}\right). \eqno(61)$$
The spatial extension of the corresponding eigenfunctions $(n\ge 1)$ is
so large, that they exceed a radius 70 times larger than the soliton
itself.  It is therefore questionable whether these fluctuations are
physically meaningful and should be taken into account at all for a
monopole gas of finite density.  Fortunately, this issue is not really
critical, since the contribution of such large-size fluctuations is
very close to a factor 1
$$\prod_{n=1}^{\infty} \omega_n^2 = \exp \left( {e^{-3\pi}\over
1-e^{-2\pi}}\right) \approx 1.0008. \eqno(62)$$
We are left with the contribution of the numerically found ground
state in this effective potential which---normalizing the same way as
before---reads as
$$\Delta^{-1/2} \approx {1\over \omega_0} \approx 1.03. \eqno(63)$$
Collecting all known factors together we obtain the monopole density
$$\eqalignno{\rho &= 0.0657\; {b\over a\omega_0} (g^2T)^3 \approx
0.0326\; (g^2T)^3,  &(64) \cr
\noalign{\hbox{leading to a screening mass of}}
\mu_M &\approx 0.255\; (g^2T) &(65) \cr}$$
for SU(2), which is amazingly close to the value obtained in lattice
simulations$^2$.  Of course, the contributions due to higher multipole
modes, which were neglected here (cf. eq. 35), may change this result
quantitatively.  However, the fact that the determinantal factor for
radial fluctuations is very close to unity lets us be optimistic that
the influence of those higher modes on the numerical constant in eq.
(65) will be minor.
\bigskip

\noindent {\bf 4. Conclusions}
\medskip

Finally we would like to present some numerical estimates
characteristic for QCD under conditions that may be reached in nuclear
collisions at RHIC energies.  We assume a temperature of $T$ = 300 MeV
and a coupling constant of $g$ = 2, corresponding to $\alpha_s =
g^2/4\pi \approx 0.32$.  We note that while this value of $\alpha_s$
would justify a perturbative QCD approach, the monopole charge $g^{-1}$
= 0.5 may justify the neglect of monopole-monopole interactions.  The
magnetic screening mass for SU(3) we obtain using the above values is
$\mu_M \approx$ 375 MeV yielding a screening length of $\mu_M$ =
0.53 fm for static magnetic fields.

The size and the total energy of an average monopole-soliton can also
be estimated using the scale representation of eq. (46):
$$\rho_M = \int_0^{\infty} d\lambda\; n(\lambda)\; e^{-\lambda
a\tilde{\beta}} \eqno(66)$$
with $n(\lambda) = n_0\lambda^3$.
The average soliton size is obtained as
$$R_M = \langle \lambda^{-1} \rangle =
{1\over 3} a\tilde{\beta} \approx {6.10\over g^2T},\eqno(67)$$
which is about 1 fm at this temperature.  Unfortunately this exceeds
the average distance $d_0 \approx 4/g^2T$ between monopoles at the
equilibrium density (64) indicating a possible breakdown of the dilute
gas approximation employed here.  On the other hand, the average
interaction energy between monopoles
$$E_{\hbox{\eightrm int}} = (4\pi g^2d_0)^{-1} \approx 0.02\; T
\eqno(68)$$
is quite small compared with the average monopole mass
$$\langle E\rangle = E[u_0]/g^2 = 4\; T. \eqno(69)$$
This indicates that higher-order terms in the effective action $S_3[A]$
for superpositions of monopoles are small, and gives reason to believe
that the dilute gas approximation may be quite trustworthy. It allows
for an intuitive understanding of the mechanism for static magnetic
screening in thermal gauge theories. We hope to return to the
problem of evaluating the determinant of eigenvalues of higher modes,
as well as to the question of non-static loop corrections
in the future.
\bigskip

\noindent {\bf Acknowledgements}
\medskip

We appreciate discussions with S. Gavin and M. Thoma at the
beginning of this work.  We thank M. Baker, M. Gyulassy, and J. Polonyi
for useful remarks concerning a preliminary version of our manuscript.
T.S.B. acknowledges the support from the U.S. Department of Energy
during his visit at Duke University (grant DE-FG05-90ER40592).
\vfill\eject

\noindent {\bf References}
\bigskip

{\frenchspacing
\item{1.} A. D. Linde, {\sl Phys. Lett. {\bf 96 B}}, 289 (1980);
D. J. Gross, R. D. Pisarski, and L. G. Yaffe, {\sl Rev. Mod. Phys. {\bf
53}}, 43 (1981).

\item{2.} A. Billoire, G. Lazarides, and Q. Shafi, {\sl Phys. Lett.
{\bf 103B}}, 450 (1981);
T. A. DeGrand and D. Toussaint, {\sl Phys. Rev. {\bf D25}}, 526 (1982).

\item{3.} R. D. Pisarski, {\sl Phys. Rev. Lett. {\bf 63}}, 1129 (1989).

\item{4.} O. K. Kalashnikov, {\sl Pis'ma Zh. Eksp. Teor. Fiz. {\bf
39}}, 337 (1984) [{\sl JETP Lett. {\bf 39}}, 405 (1984)]; {\sl Phys.
Lett. {\bf B279}}, 367 (1992).

\item{5.} M. Baker, J. S. Ball, F. Zachariasen, CERN-TH6445/92.

\item{6.} E. Braaten and R. D. Pisarski, {\sl Phys. Rev. {\bf D42}},
2156 (1990).

\item{7.} B. M\"uller and A. Trayanov, {\sl Phys. Rev. Lett. {\bf
68}}, 3387 (1992);
C. Gong, preprint DUKE-TH-92-39, to appear in {\sl Phys. Lett. B}.

\item{8.} N. P. Landsman, {\sl Nucl. Phys. {\bf B322}}, 498 (1989).

\item{9.} M. Oleszczuk and J. Polonyi, {\sl Nucl. Phys. {\bf A544}},
523c (1992).

\item{10.} See also the derivation of the dilute gas approximation
for instantons in: C. G. Callan, Jr., R. Dashen, and D. J. Gross,
{\sl Phys. Rev. {\bf D17}}, 2717 (1978).

\item{11.} A. M. Polyakov, {\sl Nucl. Phys. {\bf B120}}, 429 (1977).

\item{12.} G. t'Hooft, {\sl Nucl. Phys. {\bf 379}}, 276 (1974); A. M.
Polyakov, {\sl Pis'ma Zh. Eksp. Teor. Fiz. {\bf 20}}, 430 (1974);
[{\sl JETP Lett. {\bf 20}}, 194 (1974)].

\item{13.} T. T. Wu and C. N. Yang, in: {\sl Properties of Matter under
Unusual Conditions}, ed. H. Mark and S. Fernbach (Interscience, New
York, 1969).

\item{14.} T. Yoneya, {\sl Phys. Rev. {\bf D16}}, 2567 (1977);

R. A. Brandt and F. Neri, {\sl Nucl. Phys. {\bf B161}}, 253 (1979).

\item{15.} H. Arod\'z, {\sl Phys. Rev. {\bf D27}}, 1903 (1983).

\item{16.} T. Akiba, H. Kikuchi and T. Yanagida, {\sl Phys. Rev. {\bf
D40}}, 588 (1989).

\item{17.} E. Witten, {\sl Phys. Rev. Lett. {\bf 38}}, 121 (1977).

\item{18.} P. Arnold and L. McLerran, {\sl Phys. Rev. {\bf D36}}, 581
(1987).

\item{19.} L. Carson and L. McLerran, {\sl Phys. Rev. {\bf D41}}, 647
(1990); L. Carson, X. Li, L. McLerran, and R. T. Wang, {\sl Phys. Rev.
{\bf D42}}, 2127 (1990).

\item{20.} J. L. Gervais and B. Sakita, {\sl Phys. Rev. {\bf D11}},
2943 (1975); E. Tomboulis, {\sl Phys. Rev. {\bf D12}}, 1678 (1975).
\vfill}\eject

\noindent {\bf Appendix A}
\bigskip

In this appendix we derive the color charge density (eq.6) from the dilute
gas partition function (4) in the weak external field limit.
{}From eq.(4) we obtain the following expression for the logarithm of the
partition function
$$ \ln Z = - \sum_{\omega} \sum_{q_3} \ln \left( 1 - e^{-\beta \omega}
e^{\Theta q_3} \right), \eqno(A.1)$$
which after factorizing the color charge independent one-gluon partition
functions can be casted into the form
$$ \ln Z = - \sum_{\omega} \sum_{q_3} \left[ \ln \left( 1 - e^{- \beta \omega}
\right) \; + \; \ln \left( 1 + n(\omega) \left(1-e^{\Theta q_3} \right) \right)
\right], \eqno(A.2)$$
where
$$ n(\omega) = {1 \over e^{\beta \omega}-1} \eqno(A.3) $$
is the Bose-Einstein distribution function.

Restoring the generality of the discussion we replace now $\Theta q_3$ again
by $\beta \phi^a \hat Q^a$ and write the summation over $q_3$ as the adjoint
color trace $tr$. We arrive at
$$ \ln Z = - \sum_{\omega} \left[ \ln \left(1-e^{-\beta \omega} \right) tr 1 \;
+ \; tr \ln \left(1+n(\omega) \left(1-e^{\beta \phi^a \hat Q^a} \right) \right)
\right]. \eqno(A.4)$$
Separating now the external field independent term,
$$ \ln Z_0 = -(N^2-1) \sum_{\omega} \ln \left(1-e^{-\beta \omega} \right),
\eqno(A.5)$$
we are left with
$$ \ln Z = \ln Z_0 - \sum_{\omega} tr \ln \left( 1+n(\omega) \left(1-e^{\beta
\phi^a \hat Q^a} \right) \right), \eqno(A.6)$$
which can be approximated in the dilute limit $n(\omega) \ll 1$ by
$$ \ln Z \approx \ln Z_0 - \sum_{\omega} n(\omega) tr \left(1-e^{\beta \phi^a
\hat Q^a} \right). \eqno(A.7)$$
Finally we use the fact that the external field $\phi^a$ is infinitesimal
in case of seeking for a linear response of the gluon medium. We obtain
$$\ln Z \approx \ln Z_0 + \sum_{\omega} n(\omega) \; tr \left( \beta \phi^a
\hat Q^a + {\beta^2 \over 2}\phi^a \phi^b \hat Q^a \hat Q^b + ... \right),
\eqno(A.8)$$
which, normalizing the color magnetic charge generators according to the
adjoint representation of the SU(N) algebra,
$$ tr \left( \hat Q^a \hat Q^b \right) = - {N \over g^2} \delta^{ab},
\eqno(A.9)$$
leads to
$$ \ln Z \approx \ln Z_0 - {\beta^2 N \over 2g^2} \phi^a \phi^a \sum_{\omega}
n(\omega). \eqno(A.10)$$
The definition (5) of the color charge density leads finally to
$$\rho_E^a = {1 \over \beta V}{\partial \over \partial \phi^a} \ln Z
\approx - {N \over Vg^2T} \phi^a \sum_{\omega} n(\omega), \eqno(A.11)$$
which is equivalent to eqs. (6) and (7).

To calculate the screening of static electric fields, one replaces the
scalar magnetic potential by a Coulomb potential
$$E_i^a = - \partial_i \phi_E^a, \eqno(A.12)$$
and the magnetic charge operator $\hat Q^a$ by the electric charge operator
$\hat Q_E^a$. The sole change in the derivation following equation (A.1)
then is that (A.9) is now replaced by
$$tr \left(\hat Q_E^a \hat Q_E^b \right) = - Ng^2 \delta^{ab}. \eqno(A.13)$$
We finally obtain for the induced color-electric charge density:
$$ \rho_E^a \approx - {N g^2 \over V T} \phi_E^a \sum_{\omega} n(\omega).
\eqno(A.14)$$
Using the free gluon dispersion relation we have
$$\sum_{\omega} n(\omega) = V \int {d^3k \over (2\pi)^3}
\left( e^{\beta|k|} -1\right)^{-1} = {1\over 3} VT^3, \eqno(A.15)$$
hence
$$\rho_E^a = -{N\over 3} g^2T^2 \phi^a \equiv - \mu_E^2 \phi^a.
\eqno(A.16)$$
This is the standard perturbative result for the static elctric
screening mass in thermal Yang-Mills theories.
\vfill\eject

\noindent {\bf Appendix B}
\bigskip

In this appendix we derive the expressions (21-24) for the magnetic field
and energy from the ansatz (16) for the vector potential.  It is useful
here to introduce some short-hand notations.  We write
$$A_{ia} = A^+P_{ia}^+ + A^-P_{ia}^- + A^0P_{ia}^0. \eqno(B.1)$$
For calculating derivatives of this expression we start with the
projectors.  First we note that
$$\partial_j n_i = {1\over r}(\delta_{ij} - u_iu_j)
\eqno(B.2)$$
and
$$\delta_j(n_in_k) = {1\over r} (\delta_{ij},n_k
+\delta_{jk}n_i - 2n_in_kn_j). \eqno(B.3)$$
It is easy to obtain then
$$\epsilon_{ijk} \partial_j (\epsilon_{ka\ell} n_{\ell}) =
{1\over r} (\delta_{ia} + n_in_a) \eqno(B.4)$$
and
$$\epsilon_{ijk} \partial_j (n_kn_a) = {1\over r}
\epsilon_{iak} n_k. \eqno(B.5)$$
Using this we obtain the following expressions for the curls of
the projectors
$$\eqalignno{\epsilon_{ijk}\partial_j P_{ka}^+ &= -{1\over 2r}
\epsilon_{iak}n_k + {i\over 2r} (\delta_{ia}+u_in_a) =
{i\over r}\left( P_{ia}^+ +P_{ia}^0\right), &(B.6) \cr
\epsilon_{ijk}\partial_j P_{ka}^- &= -{1\over 2r} \epsilon_{iak} n_k
- {i\over 2r} (\delta_{ia} +n_in_a) =
{i\over r} \left(P_{ia}^- + P_{ia}^0\right) &(B.7) \cr
\noalign{\hbox{and}}
\epsilon_{ijk}\partial_jP_{ka}^0 &= {1\over r} \epsilon_{iak} n_k =
-{i\over r} \left( P_{ia}^+ -P_{ia}^-\right). &(B.8) \cr}$$
It is now easy to decompose the curl of the vector potential
remembering that $\partial_j$ acts on pure radial functions
as $n_j {d\over dr}$.  Using
$$\eqalignno{ \epsilon_{ijk}n_j P_{ka}^+ &= -{1\over 2}
\epsilon_{iaj} n_j + {i\over 2} (\delta_{ia} - n_in_a) =
iP_{ia}^+, &(B.9)\cr
\epsilon_{ijk}n_j P_{ka}^- &= -iP_{ia}^-, &(B.10)\cr
\noalign{\hbox{and}}
\epsilon_{ijk}n_j P_{ka}^0 &= 0 &(B.11) \cr
\noalign{\hbox{we get}}
\epsilon_{ijk}\partial_j A_{ka} &= iP_{ia}^+ \left(
{dA^+\over dr} + {1\over r} A^+ - {1\over r} A^0\right) + c.c +
iP_{ia}^0 \left( {1\over r} A^+ - {1\over r} A^-\right). &(B.12)
\cr}$$
We note that $(A^+)^* = A^-,\; (P_{ia}^+)^* = P_{ia}^-$.

After evaluating the abelian part of the magnetic field (eq. 21) we
turn to the calculation of the nonabelian part
$$N_{ia} = {1\over 2} \epsilon_{abc} \epsilon_{ijk} A_{jb}
A_{kc}. \eqno(B.13)$$
Noting that
$$\epsilon_{abc}\epsilon_{ijk} = \det \left\vert
\matrix{\delta_{ai}\quad \delta_{aj}\quad \delta_{ak}\cr
\delta_{bi}\quad \delta_{bj}\quad \delta_{bk} \cr
\delta_{ci}\quad \delta_{cj} \quad \delta_{ck}\cr}\right\vert
\eqno(B.14)$$
we obtain
$$N_{ia} = \textstyle{{1\over 2}} \delta_{ai} \left( A_{jj} A_{kk} -
A_{jj}^2\right) - A_{kk} A_{ai} + A_{ai}^2. \eqno(B.15)$$
Now it is straightforward to evaluate $A_{ai}^2$ and its traces,
using the projectors.  We get
$$\eqalignno{ A_{jj} &= A^+ + A^- + A^0, &(B.16) \cr
(A^2)_{ai} &= A^+A^+P_{ai}^+ + A^-A^-P_{ai}^- + A^0A^0P_{ai}^0
&(B.17)\cr
\noalign{\hbox{and}}
(A^2)_{jj} &= (A^+)^2 + (A^-)^2 + (A^0)^2. \cr}$$
Inserting these expressions into (B.15) we get
$$N_{ia} = \delta_{ai} (A^+A^-+A^-A^0+A^0A^+) - A^+(A^-+A^0) P_{ai}^+
- A^-(A^++A^0) P_{ai}^0 - A^0(A^++A^-)P_{ai}^0. \eqno(B.18)$$
Noting that $P_{ai}^+ = P_{ia}^-,\; P_{ai}^- = P_{ia}^+$ and
$\delta_{ia} = P_{ia}^+ + P_{ia}^- + P_{ia}^0$, we finally get the
simple result
$$N_{ia} = A^+A^0P_{ia}^+ + A^-A^0P_{ia}^- + A^+A^-P_{ia}^0.
\eqno(B.19)$$
The definition of the magnetic field can now be used to
obtain the decomposition
$$\eqalignno{B_{ia} &= B^+P_{ia}^+ + B^-P_{ia}^- + B^0P_{ia}^0.
&(B.20) \cr
\noalign{\hbox{We get}}
B^+ &= i {d\over dr} A^+ + {i\over r}A^+ - {i\over r} A^0
-A^+A^0, &(B.21) \cr
B^- &= -i {d\over dr} A^- - {i\over r} A^- + {i\over r}A^0 -
A^-A^0 &(B.22)\cr
\noalign{\hbox{and}}
B^0 &= {i\over r} A^+ - {i\over r} A^- -A^+A^-. &(B.23) \cr}$$

Now we include the trivial factor ${1\over r}$ in the definition of
the vector potential using the notation $A^+={1\over r}a^+$, etc.  We
get
$$\eqalignno{ B^+ &= {1\over r} \left( i {d\over dr}a^+ -{1\over r}a^0
(i+a^+)\right), &(B.24) \cr
B^- &= {1\over r} \left( -i {d\over dr} a^- +{1\over r}a^0(i-a^-)\right)
&(B.25) \cr
\noalign{\hbox{and}}
B^0 &={1\over r^2} (ia^+-ia^--a^+a^-). &(B.26)\cr}$$

One realizes at this point that the form of the ansatz (16) is simple
in terms of the magnetic field giving
$$\eqalignno{a^++i &= ue^{i\phi}, &(B.27) \cr
a^- - i &= ue^{-i\phi}, &(B.28) \cr
a^0 &= w. &(B.29) \cr}$$
Using this we arrive at
$$\eqalignno{ a^+-a^- &= 2iu\sin\phi-2i, &(B.30) \cr
a^+a^- &= u^2 + iue^{i\phi} - iue^{-i\phi} + 1 = 1-2u\sin\phi + u^2
&(B.31) \cr
\noalign{\hbox{whence finally we obtain}}
B^+ &= {1\over r} \left( i {du\over dr}\; e^{i\phi} -
u{d\phi\over dr}\; e^{i\phi} -{1\over r} wue^{i\phi}\right), &(B.32) \cr
B^- &= {1\over r} \left( -i {du\over dr} \; e^{-i\phi} -
u{d\phi\over dr}\; e^{-i\phi} - {1\over r} wue^{-i\phi}\right) &(B.33) \cr
\noalign{\hbox{and}}
B^0 &= {1\over r^2} (1-u^2). &(B.34) \cr}$$
To obtain the magnetic flux and the magnetic charge is easy observing
that $P_{ia}^+$ and $P_{ia}^-$ are orthogonal to $n_i$.  So we arrive
at eq. (22)
$$n_iB_{ia} = n_a(1-u^2). \eqno(B.35)$$
In order to calculate the energy density we note that since $B^- =
(B^+)^*,\; P_{ia}^-= (P_{ia}^+)^*$,
$$\eqalignno{{\cal E} &= \textstyle{{1\over 2}} B_{ia} B_{ia} =
B^+\cdot B^- + \textstyle{{1\over 2}} B^0\cdot B^0 &(B.36)\cr
\noalign{\hbox{is whence}}
{\cal E} &= \vert B^+\vert^2 + {1\over 2} B^0B^0 =
{1\over r^2} \left\vert i{du\over dr} - u
\left( {d\phi\over dr} + {w\over r}\right)
\right\vert^2 + {1\over 2r^4} (1-u^2)^2. &(B.37) \cr}$$
Finally it gives rise to the following reduced dimensional action
$$S_3 = {1\over g^2T} E = {4\pi\over g^2T} \int_0^{\infty} dr \left[
\left( {du\over dr}\right)^2 + u^2 \left({d\phi\over dr} + {w\over
r}\right)^2 + {1\over 2r^2} (1-u^2)^2\right] \eqno(B.38)$$
as presented in eq. (24).

It remains to calculate the action of an infinitesimal gauge
transformation on the monopole ansatz.  From the general expression
$$\delta A_{ia} = {\cal D}_{iab} \Lambda_b \eqno(B.39)$$
we conclude that $\Lambda_b = n_b\cdot\Lambda(r)$ is the only form not
leaving the ansatz's configuration space, because only the derivative
of $n_i$ is expressable through a combination of $P_{ij}^{\pm}$ and
$P_{ij}^0$.  Noting that
$$\eqalignno{D_{iab} &= \delta_{ab} \partial_i - \epsilon_{acb} A_{ic}
&(B.40) \cr
\noalign{\hbox{and}}
\epsilon_{acb}n_b &= -iP_{ac}^+ + iP_{ac}^- &(B.41) \cr
\noalign{\hbox{we obtain}}
D_{iab}n_b\Lambda &= n_in_a{d\Lambda\over dr} + {1\over r}
(\delta_{ia}-n_in_a)\Lambda - i(P_{ac}^+-P_{ac}^-) A_{ci} &(B.42)\cr}$$
which, upon using the generic form (A.1) for the vectorpotential,
yields
$$\eqalignno{\delta A^+ &= {1\over r}\Lambda - iA^+\cdot\Lambda,
&(B.43)\cr
\delta A^- &= {1\over r} \Lambda + iA^-\cdot\Lambda &(B.44) \cr
\noalign{\hbox{and}}
\delta A^0 &= {d\over dr}\Lambda. \cr}$$
Replacing now the expressions for $A^{\pm}$ and $A^0$ we finally
arrive at
$$\eqalignno{\delta u&= 0, &(B.45) \cr
\delta\phi &= -\Lambda &(B.46) \cr
\noalign{\hbox{and}}
\delta w &= r {d\over dr}\Lambda, &(B.47) \cr}$$
as claimed in eqs. (28).
\vfill\eject

\noindent {\bf Appendix C}
\bigskip

In this appendix we briefly describe the methods we employed to obtain
the numerical results of this paper.

The classical monopole-soliton as a solution of eq. (43) can not be
obtained by direct numerical integration, because the fixed boundary
condition $u(\infty) =0, u(0) =1$.  Instead we applied the
energy-functional relaxation updating the field $u_n = u_0(x)$ known
on a lattice of $x_n = \lambda\cdot n\cdot\Delta r = n \cdot\Delta x$
points due to the conjugate gradient method
$$u'_n = u_n -\epsilon\; {\delta E[u]\over \delta u_n} \eqno(C.1)$$
using $\epsilon$ = 0.01, $\Delta x$ = 0.01 - 0.05 and 200 - 1000 grid
points.  We obtained the same solution by a second order conjugate
gradient method.

We also found the eigenvalues of the effective potential (59) for
monopole-like fluctuations numerically.  The first method employed a
``shooting'' algorithm, integrating from $x=0$ and from $x=\infty$
up to a matching point, and modifying the initial derivatives in order
to fit the logarithmic derivative at the matching.  The second method
directly located the zeros of the determinant of the discretized
matrix
$$\Omega^2_{mn}-\omega^2 \Delta x^2 \delta_{mn} = -\delta_{n,m+1} -
\delta_{n,m-1} + \delta_{mn}
\left( 2 +\Delta x^2 (V_{\hbox{\sevenrm eff}} [U_n,n]-\omega^2)\right)
\eqno(C.2)$$
in the interval $0 \le \omega \le 1$.
The evaluation of the determinant made use of the simple recursion
formula for tri-diagonal $N\times N$ matrices
$$D_N = q_N\cdot D_{N-1} - D_{N-2}, \eqno(C.3)$$
where $q_N$ is the $N$-dependent diagonal element and $D_0=1$,
$D_1=q_1$ start the recurrence.
\vfill\eject

\noindent {\bf Figure Captions}
\bigskip

\itemitem{Fig. 1} The classical magnetic monopole soliton $u_0(r)$
which minimizes the energy functional $E[u] + \Delta E[u]$
is shown as solid line.  The resulting monopole charge
contained inside a sphere of radius $r$ is represented by the dashed line.
\bigskip

\itemitem{Fig. 2} The effective potential for radial magnetic monopole
fluctuations $\delta u(r)$ is plotted as a function of the scaled radius
$x=\lambda r$.  The ground state energy $\omega_0^2=0.95$ in this potential
is indicated by a horizontal straight line.

\end